\documentclass[aps,prl,twocolumn,showpacs,amsmath,amssymb]{revtex4-1}
\usepackage{amsmath}
\usepackage{graphicx}
\usepackage{subfigure}
\usepackage{epstopdf}
\usepackage{color}
\usepackage{multirow}
\usepackage{setspace}
\usepackage{overpic}
\usepackage{amssymb}
\usepackage[dvipdfmx, bookmarksnumbered, pdfstartview=FitH,colorlinks,urlcolor=blue, citecolor=blue,linkcolor=blue] {hyperref}
\usepackage{lineno}
\usepackage{bm}
\usepackage{rotating}
\usepackage[utf8]{inputenc}
\let\oldequation\equation
\let\oldendequation\endequation

\renewenvironment{equation}
  {\linenomathNonumbers\oldequation}
  {\oldendequation\endlinenomath}

\begin{document}

\title{\bf \boldmath
Observation of the Decay $D^0\to \rho^-\mu^+\nu_\mu$}

\author{
M.~Ablikim$^{1}$, M.~N.~Achasov$^{10,c}$, P.~Adlarson$^{67}$, S. ~Ahmed$^{15}$, M.~Albrecht$^{4}$, R.~Aliberti$^{28}$, A.~Amoroso$^{66A,66C}$, M.~R.~An$^{32}$, Q.~An$^{63,49}$, X.~H.~Bai$^{57}$
% ...
\\
\vspace{0.2cm}
(BESIII Collaboration)\\
\vspace{0.2cm} {\it
$^{1}$ Institute of High Energy Physics, Beijing 100049, People's Republic of China\\
% ...
$^{4}$ Bochum Ruhr-University, D-44780 Bochum, Germany\\
% ...
$^{10}$ G.I. Budker Institute of Nuclear Physics SB RAS (BINP), Novosibirsk 630090, Russia\\
% ...
$^{15}$ Helmholtz Institute Mainz, Staudinger Weg 18, D-55099 Mainz, Germany\\
% ...
$^{28}$ Johannes Gutenberg University of Mainz, Johann-Joachim-Becher-Weg 45, D-55099 Mainz, Germany\\
% ...
$^{32}$ Liaoning Normal University, Dalian 116029, People's Republic of China\\
% ...
$^{49}$ State Key Laboratory of Particle Detection and Electronics, Beijing 100049, Hefei 230026, People's Republic of China\\
% ...
$^{57}$ University of Jinan, Jinan 250022, People's Republic of China\\
% ...
$^{63}$ University of Science and Technology of China, Hefei 230026, People's Republic of China\\
% ...
$^{66}$ University of Turin and INFN, (A)University of Turin, I-10125, Turin, Italy; (B)University of Eastern Piedmont, I-15121, Alessandria, Italy; (C)INFN, I-10125, Turin, Italy\\
$^{67}$ Uppsala University, Box 516, SE-75120 Uppsala, Sweden\\
% ...
$^{c}$ Also at the Novosibirsk State University, Novosibirsk, 630090, Russia\\
% ...
}
}

\begin{abstract}
By analyzing an $e^+e^-$ annihilation data sample corresponding to an integrated luminosity
of $2.93~\mathrm{fb}^{-1}$ collected at a center-of-mass energy of
3.773 GeV with the BESIII detector, we measure the branching fraction of the $D^0\to \rho^- \mu^+\nu_\mu$
decay for the first time.  We obtain ${\mathcal B}_{D^0\to \rho^- \mu^+\nu_\mu}=(1.35\pm0.09_{\rm stat}\pm0.09_{\rm syst})\times 10^{-3}$.
Using the world average of ${\mathcal B}_{D^0\to \rho^- e^+\nu_e}$, we find a branching fraction ratio of
${\mathcal B}_{D^0\to \rho^- \mu^+\nu_\mu}/{\mathcal B}_{D^0\to \rho^- e^+\nu_e}=0.90\pm0.11$, which
agrees with the theoretical expectation of lepton flavor
universality within the uncertainty.
Combining the world average of ${\mathcal B}_{D^+\to \rho^0 \mu^+ \nu_\mu}$ and the lifetimes of $D^{0(+)}$,
we obtain a partial decay width ratio of ${\Gamma}_{D^0\to \rho^- \mu^+ \nu_{\mu}}/(2{\Gamma}_{D^+\to \rho^0 \mu^+ \nu_{\mu}}) = 0.71\pm0.14$,
which is consistent with the isospin symmetry expectation of unity within $2.1\sigma$. For the reported values of ${\mathcal B}_{D^0\to \rho^- \mu^+\nu_\mu}/{\mathcal B}_{D^0\to \rho^- e^+\nu_e}$ and ${\Gamma}_{D^0\to \rho^- \mu^+ \nu_{\mu}}/2{\Gamma}_{D^+\to \rho^0 \mu^+ \nu_{\mu}}$, the uncertainty is the quadratic sum of the statistical and systematic uncertainties.
\end{abstract}


\maketitle

\oddsidemargin  -0.2cm
\evensidemargin -0.2cm

Lepton flavor universality~(LFU) is usually thought of as a basic property of the Standard Model (SM)~\cite{Salam1964,Fajfer2012,Fajfer2015,Guo2017}.
It postulates that the couplings between the three families of leptons and gauge bosons do not depend on the lepton flavor.
Experimental studies of semileptonic decays of pseudoscalar mesons are important to test LFU and explore possible new physics.
Since 2012, tests of LFU have been carried out in several semileptonic $B$ decays at BaBar, Belle, and LHCb.
The measured branching fraction ratios
${\mathcal R}^{\bar D^{(*)}}_{\tau/\ell}={\mathcal B}_{B\to \bar
  D^{(*)}\tau^+\nu_\tau}/{\mathcal B}_{B\to \bar
  D^{(*)}\ell^+\nu_\ell}$~($\ell=\mu$, $e$)~\cite{babar_1,babar_2,lhcb_1,belle2015,belle2016,lhcb_1a,belle2019}
indicate a $3.1\sigma$ deviation from the value predicted in the SM~\cite{hflav2018}.
This tension stimulated development of various theoretical models~\cite{BFajfer2012,Fajfer2012,Celis2013,Crivellin2015,Crivellin2016,Bauer2016}.
In this context, investigations of exclusive semileptonic $D$ decays give important complementary tests of LFU.
In recent years, BESIII reported tests of $\mu{\text -}e$ LFU with the semileptonic decays $D\to X \ell^+\nu_\ell$ ($X=\bar K$, $\pi$, $\omega$, and $\eta$)~\cite{epjc76,bes3-D0-Kmuv,bes3-pimuv,bes3-omegamuv,bes3-etamuv}.
For each decay, the difference between the measured branching fraction ratio
(${\mathcal R}^{X}_{\mu/e}={\mathcal B}_{D\to X \mu^+\nu_\mu}/{\mathcal B}_{D\to X e^+\nu_e}$) and the corresponding SM prediction is less than $1.7\sigma$. The decay $D^0\to\rho^-\mu^+\nu_\mu$,
calculated using the quark potential model in 1989~\cite{isgw}, has not yet been measured.
Observation of this decay and verification of the SM prediction for ${\mathcal R}_{\mu/e}^{\rho^-}$ offer a crucial LFU test.

In addition to the quark potential model work~\cite{isgw}, the branching fraction of $D^0\to\rho^- \mu^+\nu_\mu$ has been calculated using
QCD light-cone sum rules (LCSR)~\cite{Wuyl,ly}, the light-front quark model (LFQM)~\cite{cheng},
the covariant confined quark model (CCQM)~\cite{soni1,soni2},
the chiral unitarity approach ($\chi$UA)~\cite{Sekihara},
and the relativistic quark model~(RQM)~\cite{rqm}.
The predicted branching fractions are in the range of $(1.55 - 2.01)\times10^{-3}$.
This decay also provides an opportunity to determine the $c\to d$ Cabibbo-Kobayashi-Maskawa (CKM) matrix element $|V_{cd}|$.  Furthermore, the measured branching fraction helps constrain lattice QCD calculations on the hadronic form factors of semileptonic $D$ and $B$ decays.
More precise calculations of branching fractions and hadronic form factors are key inputs in the determination of CKM parameters~\cite{Koponen1,Koponen2,Brambilla,Bailey} which allow important tests of CKM matrix unitarity.

Under the assumption of isospin symmetry, the partial width ratio ${\mathcal R}^{\rho,\ell}_{\rm IS}=\Gamma_{D^0\to \rho^-\ell^+\nu_\mu}/2\Gamma_{D^+\to \rho^0\ell^+\nu_\mu} =({\mathcal B}_{D^0\to \rho^-\ell^+
\nu_\mu}\cdot \tau_{D^+})/(2{\mathcal B}_{D^+\to \rho^0\ell^+\nu_\mu}\cdot \tau_{D^0})$ is expected to be unity. Here, $\tau_{D^{0(+)}}$ is the lifetime of the $D^{0(+)}$ meson.
Using the world average values~\cite{pdg2020}, one obtains ${\mathcal R}^{\rho,e}_{\rm IS}=0.87\pm0.13$, which agrees with unity within the uncertainty.
A measurement of the branching fraction of the decay $D^0\to \rho^-\mu^+\nu_\mu$ allows a determination of ${\mathcal R}^{\rho,\mu}_{\rm IS}$ which tests isospin symmetry in $D^{0(+)}\to \rho^{-(0)}\mu^+\nu_\mu$ decays.

Using a data sample corresponding to an integrated
luminosity of 2.93~fb$^{-1}$~\cite{lum_bes3} taken at a
center-of-mass energy of 3.773~GeV with the BESIII detector,
we report the first observation and a branching fraction measurement of $D^0\to\rho^- \mu^+\nu_\mu$, a determination of $|V_{cd}|$
and tests of both LFU with $D^0\to\rho^- \ell^+\nu_\ell$ decays and isospin symmetry in $D^{0(+)}\to \rho^{-(0)}\mu^+\nu_\mu$ decays.
Throughout this Letter, charge conjugate channels are always implied and $\rho$ denotes the $\rho(770)$.

Details about the design and performance of the BESIII detector are given in Ref.~\cite{BESCol}.
Monte Carlo (MC) simulated data samples, produced with a {\sc geant4}-based~\cite{geant4} software package including the geometric description of the BESIII detector and the
detector response, are used to determine the detection efficiency
and to estimate the backgrounds. The simulation includes the beam-energy spread and initial-state radiation in the $e^+e^-$
annihilations modeled with the generator {\sc kkmc}~\cite{kkmc}.
The inclusive MC sample consists of the production of $D\bar{D}$
pairs with consideration of quantum coherence for all neutral $D$
modes, the non-$D\bar{D}$ decays of the $\psi(3770)$, the initial-state radiation
production of the $J/\psi$ and $\psi(3686)$ states, and the
continuum processes.
The known decay modes are modeled with {\sc
evtgen}~\cite{evtgen} using the branching fractions taken from the
Particle Data Group~\cite{pdg2020}, and the remaining unknown decays
from the charmonium states are modeled with {\sc
lundcharm}~\cite{lundcharm}. Final state radiation
from charged final state particles is incorporated with the {\sc
photos} package~\cite{photos}.
 This analysis assumes that the same form factors are applicable even in the presence of LFU violation.
The vector hadronic form factors of the semileptonic decay $D^0\to \rho^-\mu^+\nu_\mu$ are simulated with those of the $D^0\to \rho^-e^+\nu_e$ decay~\cite{bes3-rhoev},
which give good data/MC consistency.

At the center-of-mass energy of 3.773~GeV, $D^0$ and $\bar D^0$ mesons are produced in pairs without additional hadrons. This feature results in an ideal environment to study $D^0$ decays with the double-tag (DT) method.~At first,
the single-tag (ST) $\bar D^0$ meson is reconstructed using the hadronic decays $\bar D^0\to K^+\pi^-$, $K^+\pi^-\pi^0$, and
$K^+\pi^-\pi^-\pi^+$. Then, the DT candidate events, in which a $D^0\to\rho^-\mu^+\nu_\mu$ decay candidate is found in the system
recoiling against an ST $\bar D^0$ meson, are selected. The branching fraction of the $D^0\to\rho^-\mu^+\nu_\mu$ decay is determined by
\begin{equation}
\label{eq:bf}
{\mathcal B}_{D^0\to\rho^-\mu^+\nu_\mu}=N_{\mathrm{DT}}/(N_{\mathrm{ST}}^{\rm tot}\cdot \varepsilon_{D^0\to\rho^-\mu^+\nu_\mu}),
\end{equation}
where $N_{\rm ST}^{\rm tot}$ and $N_{\rm DT}$ are the yields of the ST and DT candidates in data, respectively. Here,
$\varepsilon_{D^0\to\rho^-\mu^+\nu_\mu}=\Sigma_i[(\varepsilon^i_{\rm DT}\cdot N^i_{\rm ST})/(\varepsilon^i_{\rm
  ST}\cdot N_{\rm ST}^{\rm tot})]$ is the effective signal efficiency of finding
$D^0\to\rho^-\mu^+\nu_\mu$ in the presence of the ST $\bar D^0$ meson,
where $\varepsilon_{\rm ST}$ and $\varepsilon_{\rm DT}$ are the detection efficiencies of the ST and DT candidates, respectively,
and $i$ labels the ST modes.

In this analysis, the selection criteria for $K^\pm$, $\pi^\pm$, $\gamma$, and $\pi^0$ candidates follow those employed in
Refs.~\cite{epjc76,bes3-pimuv,cpc40,bes3-Dp-K1ev,bes3-etaetapi,bes3-omegamuv,bes3-etamuv,bes3-etaX,bes3-DCS-Dp-K3pi,bes3-D-KKpipi,bes3-D-b1enu}.
For the $\bar D^0\to K^+\pi^-$ tag mode, backgrounds related to cosmic rays and Bhabha scattering events are vetoed by using the requirements described in Ref.~\cite{deltakpi}.
To distinguish the ST $\bar D^0$ mesons from combinatorial backgrounds,
we define the energy difference $\Delta E\equiv E_{\bar D^0}-E_{\mathrm{beam}}$ and the beam-constrained mass $M_{\rm
  BC}\equiv\sqrt{E_{\mathrm{beam}}^{2}{/c^4}-|\vec{p}_{\bar D^0}|^{2}{/c^2}}$,
where $E_{\mathrm{beam}}$ is the beam energy, and $E_{\bar D^0}$ and
$\vec{p}_{\bar D^0}$ are the total energy and momentum of the ST $\bar D^0$
candidate in the $e^+e^-$ center-of-mass frame, respectively.
When multiple combinations for an ST mode are present in an event,
the combination with the smallest $|\Delta E|$ per tag mode per charge
is retained for further analysis.
The ST candidates are required to be within $\Delta E\in
(-0.055,0.040)$~GeV for $\bar D^0\to K^+\pi^-\pi^0$ and $\Delta E\in
(-0.025,0.025)$~GeV for $\bar D^0\to K^+\pi^-$ and $\bar D^0\to K^+\pi^-\pi^-\pi^+$.

\begin{figure}[htbp]\centering
\includegraphics[width=1.0\linewidth]{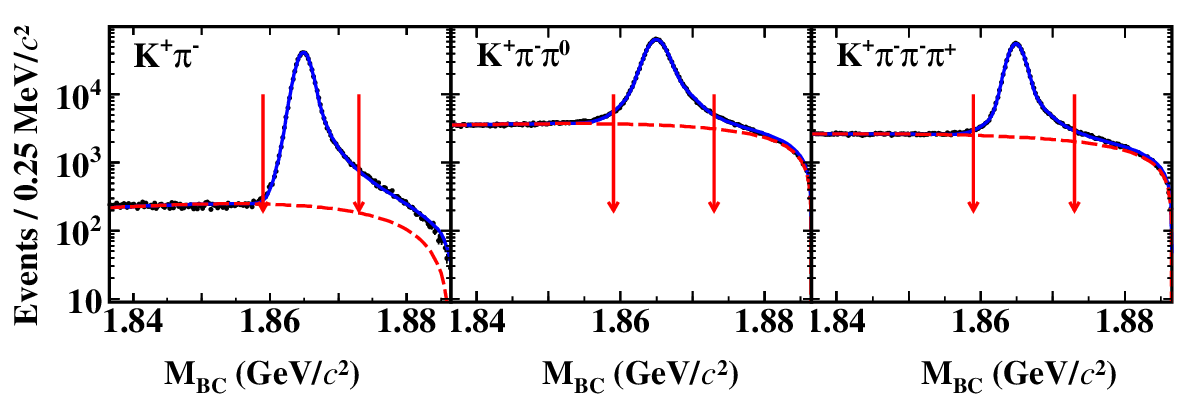}
\caption{ Fits to the $M_{\rm BC}$ distributions of the ST $\bar D^0$
  candidates.  Data are shown as dots (uncertainties are not visible at this scale).  The solid blue and
  dashed red curves are the fit results and the fitted backgrounds,
  respectively.  Pairs of red arrows indicate the $M_{\rm BC}$ selection.
}\label{fig:datafit_Massbc}
\end{figure}

Figure~\ref{fig:datafit_Massbc} shows the $M_{\rm BC}$ distributions of the accepted ST $\bar D^0$ candidates.
For each tag mode, the yield of ST $\bar D^0$ mesons is obtained from a maximum likelihood fit to
the $M_{\rm BC}$ distribution of the accepted candidates. In the fit,
the signal and background are described by the signal shape from MC simulation and
an ARGUS function~\cite{argus}, respectively.
To compensate for offsets in calibration and resolution differences between data and MC simulation, the
signal shape is convolved with a double-Gaussian function.
The means, widths and relative fractions of the Gaussian components are
free parameters in the fit.
The resulting fits to the $M_{\rm BC}$ distributions are also shown in Fig.~\ref{fig:datafit_Massbc}. Candidates in the $M_{\rm BC}$ mass window $(1.859,1.873)$ GeV/$c^2$ are kept for further analysis. For each tag mode, the yield of the ST $\bar D^0$ mesons is obtained by integrating the fitted signal shape over the $M_{\rm BC}$ mass window. The
total yield of ST $\bar D^0$ mesons is $N^{\rm tot}_{\rm
  ST}=(232.1\pm0.2_{\rm stat})\times 10^{4}$.

In the presence of the ST $\bar D^0$ mesons, candidates for $D^0\to \rho^-\mu^+\nu_\mu$ are selected from the tracks and showers which have not been used in the tag reconstruction.
The $\rho^-$ candidates are reconstructed via the $\rho^-\to \pi^-\pi^0$
decay. The selection criteria of $\pi^-$ and $\pi^0$ candidates are the same as those used in the ST selection.
The invariant mass of the $\pi^-\pi^0$ candidate is required to be within $(0.625,\,0.925)$\,GeV$/c^{2}$.
To suppress the background from hadronic $D^{0(+)}$ decays, it is required that there is no additional charged track or $\pi^0$ except for those used to form the signal and ST candidates.

The combined  information from the specific energy loss in the drift chamber, the time-of-flight system, and the electromagnetic calorimeter (EMC) is used to identify the muon candidates. The combined confidence levels for various particle hypotheses ($CL_e$, $CL_\mu$, $CL_\pi$, and $CL_K$)
are calculated. Charged tracks satisfying $CL_\mu>0.001$, $CL_\mu>CL_e$, and $CL_\mu>CL_K$ are identified as muons.
In muon identification,
no requirement of $CL_\mu>CL_\pi$ is applied because of inefficient separation between  muon and pion due to their very close masses.
Also, no muon counter information is used because most of muons in $D^0\to \rho^-\mu^+\nu_\mu$ have momenta lower than 0.6 GeV/$c$,
which are too low to leave effective information in muon counter.
To reduce misidentification of hadrons as muons, the deposited energy in the EMC of the muon candidate ($E_{\mu,\rm EMC}$) is required to be in the range
(0.125,\,0.275)\,GeV. This requirement suppresses about 40\% of total background.

The signal yield of the $D^0\to \rho^-\mu^+\nu_\mu$ decay is determined by a kinematic quantity
defined as $M^2_{\mathrm{miss}}\equiv
E^2_{\mathrm{miss}}{/c^4}-|\vec{p}_{\mathrm{miss}}|^2{/c^2}$, which is expected to
peak around zero for correctly reconstructed signal
events. Here, $E_{\mathrm{miss}}\equiv
E_{\mathrm{beam}}-E_{\rho^-}-E_{\mu^{+}}$ and
$\vec{p}_{\mathrm{miss}}\equiv
\vec{p}_{D^0}-\vec{p}_{\rho^-}-\vec{p}_{\mu^{+}}$ are the missing energy
and momentum of the DT event in the $e^+e^-$ center-of-mass frame, in
which $E_{\rho^-\,(\mu^+)}$ and $\vec{p}_{\rho^-\,(\mu^+)}$ are the energy
and momentum of the $\rho^-$\,($\mu^+$) candidates. The
$M^2_{\mathrm{miss}}$ resolution is improved using $\vec{p}_{D^0} \equiv
{-\hat{\vec{p}}_{\bar D^0}}\cdot\sqrt{E_{\mathrm{beam}}^{2}{/c^2}-m_{D^0}^{2}{c^2}}$, where
$\hat{\vec{p}}_{\bar D^0}$ is the unit vector in the momentum direction of the ST
$\bar D^0$ and $m_{D^0}$ is the $D^0$ nominal mass~\cite{pdg2020}.

The selected sample is contaminated by background events with correctly
reconstructed ST mesons but mis-reconstructed signal decays which can peak in the $M^2_{\mathrm{miss}}$ distribution.
Residual backgrounds are mainly due to misidentification between charged pion and muon.
They are dominated by few peaking background sources with a fraction of about 75\% in total.
In order to reject the peaking background from the hadronic decays $D^0\to K^0_S(\to \pi^0\pi^0)\pi^+\pi^-$ and $D^0\to K^0_S(\to \pi^+\pi^-)\pi^0(\pi^0)$, the mass recoiling against the $\bar D^0 \pi^+_{\mu\to\pi}\pi^-$ system and the invariant mass of the $\pi^+_{\mu\to\pi}\pi^-$ combination are required to be outside  $(0.458,\,0.538)$\,GeV$/c^{2}$ and $(0.468,\,0.528)$\,GeV$/c^{2}$, respectively,
where $\pi^+_{\mu\to\pi}$ denotes a track identified as a muon candidate whose mass has been replaced by the $\pi^+$ mass.
To reduce the peaking background from $D^0\to\pi^+\pi^-\pi^0$, the invariant mass of the $\rho^-\mu^+$ combination ($M_{\rho^-\mu^+}$) is required to be less than 1.5~GeV/$c^2$.
To suppress the peaking background from $D^0\to\pi^+\pi^-\pi^0\pi^0$, the maximum energy of any photon that is not used in the DT selection ($E_{\rm extra~\gamma}^{\rm max}$) is required to be less than 0.25~GeV.
With these requirements, about 88\% of $D^0\to \pi^+\pi^-\pi^0\pi^0$ background are rejected
and more than 99\% of the other backgrounds aforementioned are vetoed.
The remaining peaking background events are mainly from $D^0$ decays into $\pi^+\pi^-\pi^0\pi^0$ final states, including $D^0\to K^0_S(\to \pi^0\pi^0)\pi^+\pi^-$, $D^0\to K^0_S(\to\pi^+\pi^-)\pi^0\pi^0$, $D^0\to K^-(\to \pi^-\pi^0)\pi^+\pi^0$, and $D^0\to \pi^+\pi^-\pi^0\pi^0|_{{\rm non}{\text -}K}$. Since there is little difference in their $M^2_{\rm miss}$ shape, these four components are combined together, and will be called $D^0\to \pi^+\pi^-\pi^0\pi^0$. The remaining background events from $D^0\to K^0_S(\to \pi^+\pi^-)\pi^0$, and $D^0\to \pi^+\pi^-\pi^0$ are negligible and have been combined into the combinatorial background in further analysis.

To suppress the background from $D^0\to K^*(892)^-(\to K^-\pi^0)\mu^+\nu_\mu$,
the candidate events are further required not to be within the range $|M^2_{\mathrm{miss}\,\pi^-\to K^-}|<0.05$~GeV$^2/c^4$, where $M^2_{\mathrm{miss}\,\pi^-\to K^-}$ is the $M^2_{\rm miss}$ value calculated by replacing the mass of the charged pion candidate with the kaon mass in the calculation of $M^2_{\mathrm{miss}}$.

\begin{figure}[htbp] \centering
\includegraphics[width=1.0\linewidth]{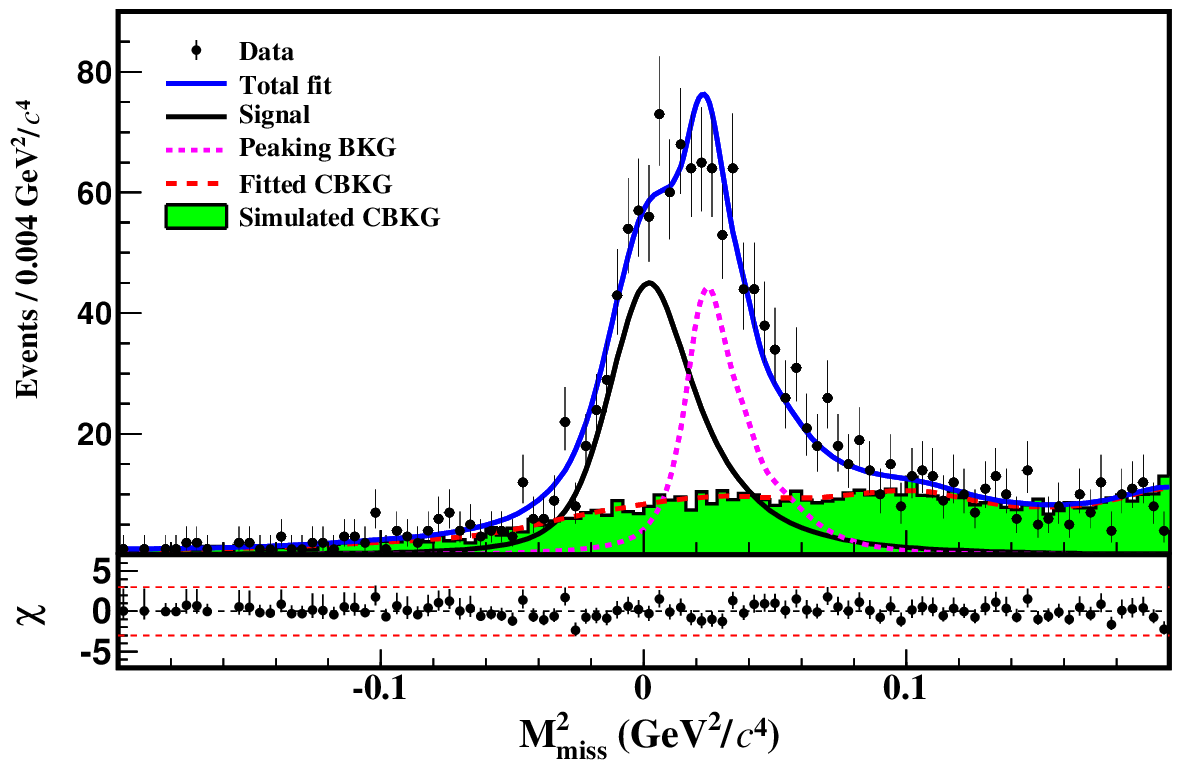}
\caption{Fit to the $M^2_{\rm miss}$ distribution of the accepted candidate events for
  $D^{+}\rightarrow\rho^-\mu^{+}\nu_{\mu}$ in data (points with error bars).  The solid blue curve
  is the fit result, the solid black curve is the semileptonic signal,
  the dashed pink curve is the peaking background (Peaking BKG) of $D^0\to \pi^+\pi^-\pi^0\pi^0$, and
  the dashed red curve is the fitted combinatorial background (Fitted CBKG).
  The filled green histogram is the simulated combinatorial background (Simulated CBKG) from inclusive MC sample.}
\label{fig:fit_Umistry1}
\end{figure}

Figure~\ref{fig:fit_Umistry1} shows the $M^2_{\mathrm{miss}}$ distribution of the accepted DT events in data. The semileptonic decay yield is
obtained from an unbinned maximum likelihood fit to the $M^2_{\mathrm{miss}}$ distribution.
In the fit, the semileptonic signal is modeled by
the MC-simulated shape convolved with a Gaussian function describing differences in resolution and calibration between data and MC simulation.
The parameters of this Gaussian function are fixed to the values obtained from a similar fit to $D^0\to \rho^-e^+\nu_e$
candidate events which have much cleaner environment and comparable momentum resolution.
The peaking background of $D^0\to \pi^+\pi^-\pi^0\pi^0$ is modeled by the $M^2_{\rm miss}$ shape derived from the
$D^0\to \pi^+\pi^-\pi^0\pi^0$ control sample in data, in which one $\pi^0$ is removed and the $\pi^+$ mass is replaced by the $\mu^+$ mass.
The non-peaking backgrounds, including the contribution from wrongly reconstructed ST candidates, are described by the MC-simulated shape obtained from the inclusive MC sample.
The yields of the signal, peaking background, and non-peaking backgrounds are free parameters in the fit. The fit result is also shown in
Fig.~\ref{fig:fit_Umistry1}. From the fit, we obtain the signal yield of $D^0\to \rho^-\mu^+\nu_\mu$ to be $N_{\rm DT}= 570\pm40_{\rm stat}$ and the yield of the peaking background of $D^0\to \pi^+\pi^-\pi^0\pi^0$ to be $373\pm36$. The statistical
significance, calculated by $\sqrt{-2{\rm ln ({\mathcal L_0}/{\mathcal
      L_{\rm max}}})}$, is greater than $10\sigma$. Here,
${\mathcal L}_{\rm max}$ and ${\mathcal L}_0$ are the maximum
likelihoods of the fits with and without the signal component, respectively, and the difference in the number of fit parameters is one.

\begin{figure}[htbp] \centering
\includegraphics[width=1.0\linewidth]{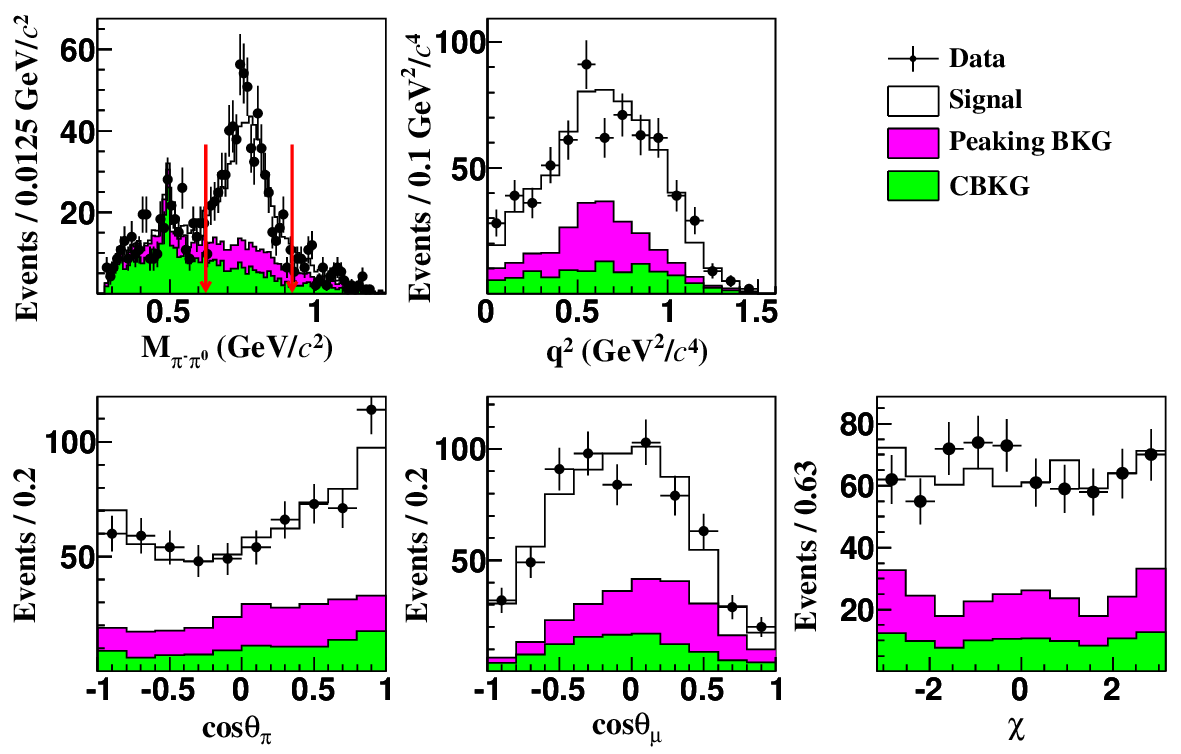}
\caption{Comparison of five kinematic variables \cite{ref:cab1, BaBar-Dp2kpienu} of the
$D^0\to \pi^-\pi^0\mu^+\nu_\mu$ candidates between data (points with error bars) and MC simulation (histograms):
the invariant mass of the $\pi^-\pi^0$ system, $M_{\pi^-\pi^0}$;
the invariant mass squared of the $\mu^+\nu_\mu$ system, $q^2$;
the angle between the momentum of the $\mu^+$ ($\pi^-$) in the $\mu^+\nu_\mu$ ($\pi^-\pi^0$) rest frame and
the momentum of the $\mu^+\nu_\mu$ ($\pi^-\pi^0$) system in the $D^0$ rest frame, $\theta_\mu$ ($\theta_\pi$);
and the angle between the normals of the decay planes defined in the $D^0$ rest frame by the $\pi^-\pi^0$ pair and the $\mu^+\nu_\mu$ pair, $\chi$.
Pink and blue histograms denote the peaking BKG and CBKG components, respectively.
Except for $M_{\pi^-\pi^0}$ to be shown,
events have been imposed with all requirements described in text and $|M^2_{\rm miss}|<0.025$~GeV$^2$/$c^4$.
In the $M_{\pi^-\pi^0}$ distribution, pair of red arrows indicate the $\rho^-$ mass window.}
\label{fig:compare}
\end{figure}

The tag-related values $N^i_{\rm ST}$, $\epsilon^i_{\rm ST}$, and $\epsilon^i_{\rm DT}$ are summarized in Table~\ref{datasignum}.
The average efficiency of detecting $D^0\to\rho^-\mu^+\nu_\mu$ decays is
$\varepsilon_{D^0\to\rho^-\mu^+\nu_\mu}=(18.22\pm0.13_{\rm stat})\%$
which
includes the branching fraction of $\pi^0\to\gamma\gamma$.
The kinematic distributions of the $D^0\to\rho^-\mu^+\nu_\mu$
candidate events agree well between data and MC simulation, as shown in Fig.~\ref{fig:compare}.

\begin{table}[htp]
\centering
\caption{The ST $\bar D^0$ yields in data ($N^i_{\rm ST}$), the ST efficiencies ($\epsilon^i_{\rm ST}$) and the DT efficiencies ($\epsilon^i_{\rm DT}$). The uncertainties are statistical only.}
\begin{tabular}{cccc}
  \hline
  $\bar D^0$ mode $i$& {$N^i_{\rm ST}$}  &{ $\epsilon^i_{\rm ST}$ (\%)} & {$\epsilon^i_{\rm DT}$ (\%)} \\ \hline
$K^+\pi^-$           & $516971  \pm    746$ & $64.28\pm0.09$ & $12.87\pm0.11$ \\
$K^+\pi^-\pi^0$      & $1099361 \pm   1327$ & $36.35\pm0.04$ & $ 6.95\pm0.08$ \\
$K^+\pi^-\pi^-\pi^+$ & $704677  \pm   1094$ & $40.26\pm0.07$ & $ 6.25\pm0.08$ \\
  \hline
\end{tabular}
\label{datasignum}
\end{table}

Inserting $N_{\rm DT}$, $\varepsilon_{D^0\to\rho^-\mu^+\nu_\mu}$, and $N_{\rm ST}^{\rm tot}$ into Eq.~(\ref{eq:bf}), we obtain
\begin{equation}
{\mathcal B}_{D^0\to\rho^- \mu^+\nu_\mu}=(1.35\pm0.09\pm0.09)\times 10^{-4},\nonumber
\end{equation}
where the first uncertainty is statistical and the second is systematic as discussed below.

In the branching fraction measurement with the DT method, most uncertainties related to the ST selection cancel.
Systematic uncertainties arise from the following sources. The uncertainty in the total yield of ST $\bar D^0$
mesons has been studied in Refs.~\cite{epjc76,cpc40,bes3-pimuv} and
is 0.5\%.
The systematic uncertainties originating from the tracking and PID efficiencies
of $\pi^\pm$ are 0.3\% and 0.2\%  per pion, respectively, based on an analysis
of DT $D\bar D$ hadronic events~\cite{D-PP}.
The muon tracking and PID efficiencies are studied
by analyzing $e^+e^-\to\gamma\mu^+\mu^-$ events.
Here, the muon identification efficiencies include the $E_{\mu,\rm EMC}$ requirement.
Using this control sample, data-MC differences are studied in the two-dimensional momentum
versus $\cos\theta$ plane. We re-weight using the obtained data-MC differences, accounting for the different distribution of events in momentum
versus $\cos\theta$ for the $D^0\to \rho^- \mu^+\nu_\mu$ signal decays.   Systematic uncertainties are obtained as the integral over the re-weighted two-dimensional distribution, giving 0.2\% and 0.2\% per muon for the muon tracking and PID efficiencies, respectively.
The uncertainty of the $\pi^0$ reconstruction is studied with DT $D\bar D$
hadronic decays of $D^0\to K^-\pi^+$, $K^-\pi^+\pi^+\pi^-$ versus $\bar
D^0\to K^+\pi^-\pi^0$, $K^0_S\pi^0$~\cite{epjc76,cpc40} and is found to be 0.6\%. The
uncertainty of the combined $E_{\rm extra~\gamma}^{\rm max}$
and $N_{\rm extra~\pi^0}$ requirements is estimated to be 1.3\% by analyzing the
DT candidate events of $D^0\to \pi^-\pi^0 e^+\nu_e$.
The uncertainty of the $M^2_{\rm miss}$ fit is found to be 6.6\% by examining the branching fraction changes with
an alternative signal shape without Gaussian smearing of the MC-simulated signal shape (0.9\%),
an MC-simulated shape of the peaking background of $D^0\to \pi^+\pi^-\pi^0\pi^0$ (5.3\%), and
combinatorial background shapes after varying the quoted branching fractions by $\pm 1\sigma$ for the two main combinatorial components of $D^0\to K^0_S\pi^+\pi^-\pi^0$ and $D^0\to K^*(892)^-\mu^+\nu_\mu$  (3.8\%).
The uncertainty arising from the finite MC statistics used to determine the efficiencies is 0.7\%.
The uncertainty due to the signal MC model is 0.3\%, determined by the difference between our nominal DT efficiency and that
determined by varying the input form factors by $\pm 1\sigma$.
Systematic uncertainties from other selection criteria are found to be negligible.
Adding these uncertainties in quadrature yields a total systematic uncertainty of 6.8\%.

\begin{table*}[htpb]
\centering
\caption{Comparison of the measured and predicted branching fractions for $D^0\to \rho^- \mu^+ \nu_{\mu}$.~The differences include both experimental and theoretical uncertainties for the LCSR and LFQM models; only experimental uncertainties are used for the other models.}
\label{table:BF}
\small
\begin{tabular}{lccccccccc}
  \hline
&BESIII& LCSR~\cite{Wuyl}& LCSR~\cite{ly}& LFQM~\cite{cheng} &CCQM~\cite{soni1}&CCQM~\cite{soni2}& $\chi$UA~\cite{Sekihara} & RQM~\cite{rqm}\\ \hline
${\mathcal B}_{D^0\to \rho^- \mu^+ \nu_{\mu}}$ ($\times 10^{-3}$) &$1.35\pm0.09\pm0.09$&$1.73^{+0.17}_{-0.13}$&$1.65\pm0.23$&$1.7\pm0.2$&2.01&1.55&1.84&1.88   \\ \hline
Difference\,($\sigma$)                                            &&2.1&1.1&1.5&5.2&1.6&3.8&4.2\\ \hline
\end{tabular}
\end{table*}

In summary, the semileptonic decay $D^0\to \rho^-\mu^+\nu_\mu$ has been observed for the first time. The absolute branching fraction of this decay is determined to be ${\mathcal B}_{D^0\to \rho^-
  \mu^+\nu_\mu}=(1.35\pm0.09_{\rm stat}\pm0.09_{\rm syst})\times
10^{-3}$.
Table \ref{table:BF} shows comparisons of the measured and predicted branching fractions for $D^0\to \rho^- \mu^+ \nu_{\mu}$.
Using the world average value of ${\mathcal B}_{D^0\to\rho^-
  e^+\nu_e}=(1.50\pm0.12)\times 10^{-3}$~\cite{pdg2020}, we obtain the branching fraction ratio
$R_{\mu/e}= {\mathcal B}_{D^0\to\rho^-
  \mu^+\nu_e}/{\mathcal B}_{D^0\to\rho^-
  e^+\nu_e} =0.90\pm0.11$.
This result agrees with the SM predictions $0.93-0.96$~\cite{Wuyl,Sekihara,cheng,soni1,soni2,rqm}.
Our result is consistent with LFU in $D^0\to \rho^-\ell^+\nu_\ell$ decays.
Combining the world averages of ${\mathcal B}_{D^+\to \rho^0 \mu^+ \nu_\mu}$, $\tau_{D^0}$, and $\tau_{D^+}$~\cite{pdg2020},
we determine ${\mathcal R}^{\rho,\mu}_{\rm IS}= 0.71\pm0.14$.
This ratio deviates from unity based on isospin symmetry at the level of $2.1\sigma$.
Improved measurements of $D^0\to \rho^- \mu^+ \nu_\mu$ and $D^+\to \rho^0 \mu^+ \nu_\mu$ with larger data samples~\cite{bes3-white-paper,belle2-white-paper}
in the near future will be crucial to clarify this tension.

The BESIII collaboration thanks the staff of BEPCII and the IHEP computing center for their strong support. This work is supported in part by National Key Research and Development Program of China under Contracts Nos. 2020YFA0406400, 2020YFA0406300; National Natural Science Foundation of China (NSFC) under Contracts Nos. 11775230, 11625523, 11635010, 11735014, 11822506, 11835012, 11935015, 11935016, 11935018, 11961141012; the Chinese Academy of Sciences (CAS) Large-Scale Scientific Facility Program; Joint Large-Scale Scientific Facility Funds of the NSFC and CAS under Contracts Nos. U1932102, U1732263, U1832207; CAS Key Research Program of Frontier Sciences under Contracts Nos. QYZDJ-SSW-SLH003, QYZDJ-SSW-SLH040; 100 Talents Program of CAS; INPAC and Shanghai Key Laboratory for Particle Physics and Cosmology; ERC under Contract No. 758462; European Union Horizon 2020 research and innovation programme under Contract No. Marie Sklodowska-Curie grant agreement No 894790; German Research Foundation DFG under Contracts Nos. 443159800, Collaborative Research Center CRC 1044, FOR 2359, GRK 214; Istituto Nazionale di Fisica Nucleare, Italy; Ministry of Development of Turkey under Contract No. DPT2006K-120470; National Science and Technology fund; Olle Engkvist Foundation under Contract No. 200-0605; STFC (United Kingdom); The Knut and Alice Wallenberg Foundation (Sweden) under Contract No. 2016.0157; The Royal Society, UK under Contracts Nos. DH140054, DH160214; The Swedish Research Council; U. S. Department of Energy under Contracts Nos. DE-FG02-05ER41374, DE-SC-0012069.

\end{document}